\documentclass[a4paper,12pt]{article}
\usepackage{amssymb}
\usepackage{latexsym}

\usepackage{epsfig}
\usepackage{graphicx}

\usepackage{amsmath,amsthm,amsfonts,amssymb}

\newcommand{\be}{\begin{eqnarray}}
\newcommand{\ee}{\end{eqnarray}}
\newcommand{\bea}{\begin{eqnarray}}
\newcommand{\nn}{\nonumber}
\newcommand{\eea}{\end{eqnarray}}

\newcommand{\nk}{\noindent}

\begin{document}

\begin{titlepage}
\begin{flushright}
hep-th/0103045\\ UA/NPPS-03-2001
\end{flushright}
\begin{centering}
\vspace{.8in}
{\large {\bf New Perspectives on Moving Domain Walls in
$(A)dS_{5}$ Space}}
\\

\vspace{.5in} {\bf Georgios Kofinas\footnote{gkofin@phys.uoa.gr}}
\\

\vspace{0.3in}
University of Athens\\ Physics Department\\
Nuclear and Particle Physics Section\\
Panepistimioupolis, Ilisia GR 157 71\\ Athens, Greece\\
\end{centering}

\vspace{1in}
\begin{abstract}
\nk A new moving domain wall solution is obtained for a flat
3-universe. This consists of a bulk metric depending on both time
and the extra coordinate, plus a dynamically interacting domain
wall, admitted by the metric and inhabited by the three-universe.
The matter contents are cosmological constants on the domain wall
and the bulk. The bulk space is shown to be $(A)dS_{5}$. A
remarkable fact concerning the three-universe is that its scale
factor never vanishes, even though the corresponding scale factor
of the bulk metric vanishes. The inclusion of a bulk scalar field
is discussed, neglecting back-reaction. Its normalizability and
the existence of a positive frequency or adiabatic bulk vacuum are
shown.
\\
\\
\\
\\
\\
\\
\\
\\
\\
\\
March, 2001
\end{abstract}
\end{titlepage}

\newpage

\baselineskip=18pt
\section*{1 \,\,\,Introduction}
\hspace{0.8cm} Over the last few years a lot of interest has been
raised on the possibility that our universe is a three-brane
embedded in a higher dimensional spacetime (bulk). Ordinary matter
fields are assumed to live on the brane while gravity propagates
in the whole spacetime. The major part of the work done in this
direction refers to branes sitting at a prescribed point of an
extra dimension. However, it is tempting, even inspired by
$Dp$-brane models, to consider that the 3-brane is somehow let to
move in the big spacetime. One such idea was introduced in
\cite{kehagias}, where the universe three-brane follows the
classical geodesics of a spherical background bulk geometry and
its dynamics is governed by the DBI action. A different idea,
where gravitational back-reaction effects are included in the
motion of the brane, i.e. the brane interacts with the whole bulk
metric which is not however prescribed, was analysed in
\cite{champlin,perry}.
\par
In the present paper, we will adopt the term ``moving domain
wall'' for a ($D-2$)-dimensional spacelike surface moving in a
$D$-dimensional bulk space. More explicitly, a ``domain wall that
moves'' will be a $D-2$ object (we even use the term ``moving
three-universe'' when we discuss the case $D=5$). The term
``domain wall'' will be used for a ($D-1$)-dimensional
hypersurface. When a domain wall moves in the big space, it forms
a thin shell (hypersurface). If Einstein's gravity resides in the
bulk space, the matter source carried by the moving domain wall
makes the stress-energy tensor of the domain wall a distributional
source for the Einstein equations. In such a case of a thin shell
it was proved long ago \cite{israel} that the discontinuity of the
extrinsic curvature $K_{AB}$ of the shell is related to the
energy-momentum tensor $T_{AB}$ of the matter on the shell by the
``Israel matching conditions''
\be
 [K_{AB}-K_{C}^{C}h_{AB}]=-\kappa_{(D)}^{2}T_{AB},
 \label{israel}
\ee where $h_{AB}$ is the induced metric on the shell and
$\kappa_{(D)}^{2}=8\pi G_{(D)}$. In \cite{champlin,perry} the bulk
on both sides of the wall was assumed to be static and then
consistency of the Israel conditions yielded non-trivial
relationships between the metric in the bulk and the matter on the
wall. These were proved to be compatible with the Einstein
equations in the bulk, and thus, static spacetime solutions were
obtained. Some of these solutions possess black hole or
cosmological horizons, beyond which the domain wall moves in
purely time-dependent bulks. In the present work, we follow this
general idea but we abandon the static ansatz for the $D-2$
metric, allowing for the scale factor of the metric to be a
function of both time and the extra coordinate. The matter content
of our model consists of cosmological constants on both the
(moving) domain wall and the bulk, instead of fields. The solution
found is a special solution for the flat three-universe, but is
the general solution for the zero value of some parameter
$\mathcal{C}$ encountered. In \cite{kraus} the domain wall moves
obeying the Israel conditions in a fixed black hole background. In
\cite{park} a moving domain wall solution was obtained for a
different ansatz for the bulk metric. Our analysis is performed
for $D$ dimensions in the big space, though our primary interest
lies in $D=5$, where in this case, the moving domain wall is
supposed to represent our three-universe. A remarkable feature of
our solution is that the scale factor of the three-universe never
vanishes, i.e. the universe manages to avoid collapsing, while the
bulk geometry has an apparent singularity.
\par
One of the attractive features of higher dimensional models is
that they provide new ways to solve the hierarchy problem. It has
been shown \cite{arkani} that if the higher dimensional spacetime
is approximately a product of a 4-dimensional spacetime with a
``large'' compact space, then the higher dimensional scale of
gravity, identified with the string scale, can be several orders
of magnitude lower than the effective four-dimensional Planck
scale. In a second scenario \cite{randall}, it was shown that, for
a particular four-metric depending on the bulk coordinate (non
factorizable), the Planck scale is determined by the higher
dimensional curvature, rather than the size of the extra
dimension, which may be infinite. It was also shown there, that
there is a single gravitational bound state confined to the brane,
which corresponds to the graviton. In our case, instead of an
analysis of some tensor perturbations of the metric, a study of an
additional bulk scalar field on the found background geometry
(neglecting back-reaction) has been performed. This field is seen
to be normalizable with respect to the extra coordinate, while its
time-dependent part allows a consistent definition of a bulk
vacuum.
\par
The structure of the paper is as follows: In section 2, we derive
the dynamical equations of motion of a domain wall in a broad
class of bulk metrics, generalizing in this way the result of
\cite{champlin}. Then, adopting an ansatz for the metric and for
the matter contents chosen, we arrive at some differential
relation between the bulk metric and the wall matter. In section
3, specializing to a flat three-universe, we show that the above
relation is compatible with the bulk Einstein equations and the
complete bulk solution is obtained. In section 4, the
three-universe trajectory, as well as its own time evolution is
discussed, for the various parameters of the model. In section 5,
a scalar field is considered on the found bulk background, without
back-reacting on it, and the wave equation is solved. Then,
normalizability of the field is discussed and a positive frequency
or adiabatic vacuum in the bulk space is shown to exist. Finally,
in section 6, we conclude and speculate on possible
generalizations.

\section*{2 \,\,\,Dynamic Domain Wall Motion}
\hspace{0.8cm} The purpose of this section is to derive the
equations of motion of a domain wall moving in the bulk
cosmological metric of a block-diagonal form \be
ds^{2}=g_{AB}dx^{A}dx^{B}=\varepsilon
N^{2}(t,y)dt^{2}+b^{2}(t,y)dy^{2}+g_{ij}(t,y,x)dx^{i}dx^{j},
\label{linelement}
 \ee
 where the positive definite metric $g_{ij}$
defines the line element of the ($D-2$)- dimensional space. In
this section we keep for completeness the full matrix $g_{ij}$
instead of only one scale factor for the $D-2$ part. We also keep
the symbol $\varepsilon=\pm1$ for future treatment of instanton
solutions. Throughout this article, we will adopt the following
convention for indices: capital Latin letters $A,B,...=1,...,D$
will denote full spacetime, while lower Latin $i,j,...=1,...,D-2$
are spacelike indices parallel to the moving domain wall.
\par
Let the position of the moving domain wall in the bulk space be
determined by a function $y(t)$, which we seek to find. The unit
normal to the hypersurface formed by this motion (pointing for
$Nb>0$ to the region with $y<y(t)$) is \be
n^{A}=\frac{Nb}{\sqrt{N^{2}+\varepsilon
b^{2}\dot{y}^{2}}}\,\Big(\frac{\varepsilon}{N^{2}}\,\dot{y},\,-\frac{1}{b^{2}},\,\vec{0}\,\Big),
\label{normal}
 \ee
 where $\dot{y}=\frac{dy}{dt}$. If $\tau$ is the
proper time measured by the moving domain wall, then
$\{\tau,x^{i}\}$ is a coordinate patch on the hypersurface and the
induced metric on this is written as
\be
ds_{(D-1)}^{2}=\varepsilon
d\tau^{2}+g_{ij}^{(DW)}(\tau)dx^{i}dx^{j}.
 \label{induced metric}
 \ee
 The moving domain wall proper velocity vector field is $\mathrm{e}=\frac{d}{d\tau}$.
 Obviously, $\mathrm{e}^{A}\mathrm{e}_{A}=\varepsilon$, $\mathrm{e}^{A}n_{A}=0$. The domain wall can be given in parametrized form
 by equations $t=t(\tau)$, $y=y(\tau)$. If $t(\tau)$ is to be an increasing
 function, then $\mathrm{e}^{A}$ gets the form
 \be
\mathrm{e}^{A}=\frac{1}{\sqrt{N^{2}+\varepsilon
b^{2}\dot{y}^{2}}}\,(1,\,\dot{y},\,\vec{0}\,).
 \label{proper velocity}
 \ee
 The relation between $t$ and $\tau$ is
 \be
\frac{dt}{d\tau}=\frac{1}{\sqrt{N^{2}+\varepsilon
b^{2}\dot{y}^{2}}}\,.
 \label{time-proper time}
 \ee
Obviously, the moving domain wall proper time is different than
the proper time of the bulk space.
\par
We shall now compute the extrinsic curvature
$K_{AB}=h_{A}^{C}h_{B}^{D}\nabla_{C}n_{D}$ of the hypersurface,
where $h_{AB}=g_{AB}-n_{A}n_{B}$. It is easier to do this in the
non-holonomic basis $\{\mathrm{e},n,\partial_{i}\}$. The spatial
components are
\be
K_{ij}=\frac{1}{2\sqrt{N^{2}+\varepsilon
b^{2}\dot{y}^{2}}}\left(\frac{\varepsilon
b}{N}\,\dot{y}\,\partial_{t}g_{ij}-\frac{N}{b}\,\partial_{y}g_{ij}\right).
\label{extrinsicij}
 \ee
The $00$ ($0$ refers to $\mathrm{e}$) component of $K_{AB}$ is
\be
K_{00}=\varepsilon
[\mathrm{e},n]^{0}=\frac{\varepsilon}{\sqrt{N^{2}+\varepsilon
b^{2}\dot{y}^{2}}}(N^{2}\,[\mathrm{e},n]^{t}+\varepsilon
b^{2}\dot{y}\,[\mathrm{e},n]^{y}).
 \label{extrinsic00}
 \ee
Computing in a straightforward way the Lie bracket of $\mathrm{e}$
and $n$ we obtain: \be
[\mathrm{e},n]^{t}=\frac{\varepsilon}{N^{2}+\varepsilon
b^{2}\dot{y}^{2}}\,\partial_{t}\Big(\frac{b\,\dot{y}}{N}\Big)-\frac{1}{N^{2}b}\,\partial_{y}N
 \label{commutatort}
 \ee
\be
[\mathrm{e},n]^{y}=\frac{1}{Nb^{2}}\,\partial_{t}b-\frac{\dot{y}}{N^{2}+\varepsilon
b^{2}\dot{y}^{2}}\,\partial_{y}\Big(\frac{N}{b}\Big).
 \label{commutatory}
\ee
 After some manipulation of the various terms we get the result
\be
K_{00}=\frac{1}{Nb}\frac{d}{dt}\left(\frac{b^{2}\dot{y}}{\sqrt{N^{2}+\varepsilon
b^{2}\dot{y}^{2}}}\right)-\frac{\varepsilon}{Nb}\,\partial_{y}(\sqrt{N^{2}+\varepsilon
b^{2}\dot{y}^{2}}\,),
 \label{extrinsic00final}
 \ee
 which is important for the present work. Note that when $N, b$ do
 not contain time explicitly, this result reduces to that
 obtained in \cite{champlin}.
\newline
Taking the trace of equations (\ref{extrinsicij}) we have
\be
2\sqrt{N^{2}+\varepsilon b^{2}\dot{y}^{2}}\,K=\frac{\varepsilon
b}{N}\,\frac{\partial_{t}\gamma}{\gamma}\,\dot{y}-\frac{N}{b}\,\frac{\partial_{y}\gamma}{\gamma}\,,
 \label{onescale}
 \ee
 where $K\equiv K_{ij}g^{ij}$ and $\gamma\equiv det(g_{ij})$. This is, of course, the unique component contained in
(\ref{extrinsicij}) when there exists only one scale factor in the
$D-2$ metric.
\par
We consider that the matter content of the model consists of a
cosmological constant $\Lambda_{b}$ on the domain wall and a
cosmological constant $\Lambda_{B}$ in the bulk. We shall seek
solutions in which the bulk spacetime is symmetric under
reflection in the domain wall (as in Ho$\check{r}$ava-Witten
supergravity) and thus the Israel equations (\ref{israel}) get the
totally umbilic form
\be
K_{AB}=-\frac{\Lambda_{b}}{2(D-2)}h_{AB}.
 \label{symmetric}
  \ee
We assume for the metric (\ref{linelement}) the \underline{ansatz}
\be
N=N(t)\,\,,\,\,\, b=b(t).
\label{ansatz}
 \ee
Then, the term containing the $\partial_{y}$ derivative in
(\ref{extrinsic00final}) vanishes. It is convenient to adopt the
gauge
\be
N=\frac{1}{b}\,,
\label{gauge}
\ee
which corresponds to a choice
of time. Using (\ref{symmetric}), equation
(\ref{extrinsic00final}) becomes equivalently
\be
\frac{d}{dt}\left(\frac{b^{3}\dot{y}}{\sqrt{1+\varepsilon
b^{4}\dot{y}^{2}}}\right)=c,
 \label{acceleration}
 \ee
 where
$c=-\frac{\varepsilon \Lambda_{b}}{2(D-2)}$\,\,. This can be
integrated to give
\be
\dot{y}=\pm\frac{ct}{b^{2}\sqrt{b^{2}-\varepsilon c^{2}t^{2}}}\,,
\label{motion}
 \ee
  where an irrelevant constant of integration has
been absorbed in time. Note that the $+$ sign (resp. $-$)
corresponds to $b>0$ $(<0)$. It is obvious that the $-$ sign
solutions arise from the $+$ ones, under reflection through some
plane parallel to the $(t,x^{i})$ plane.
\newline
Substituting $\dot{y}$ from (\ref{motion}) into (\ref{onescale}),
we obtain
 \be
\left(t\,\frac{\partial_{t}\gamma}{\gamma}\mp2(D-2)\right)^{2}b^{4}-\frac{1}{c^{2}}\,\Big(\frac{\partial_{y}\gamma}{\gamma}\Big)^{2}\,b^{2}+\varepsilon
t^{2}\,\Big(\frac{\partial_{y}\gamma}{\gamma}\Big)^{2}=0.
\label{algebraic} \ee
 This equation has to hold at every point visited by the domain wall. Thus, unless the domain wall remains at fixed $y$,
 i.e. $\dot{y}=0$, it has to hold over a range of $t,y$. Hence, the ansatz
(\ref{ansatz}) resulted in the above condition among the
components of the bulk metric and the cosmological constant
$\Lambda_{b}$\,. Equation (\ref{algebraic}) is quadratic in
$b^{2}$ with solution \be
b^{2}=\frac{\partial_{y}\gamma}{2c^{2}\gamma}\left(t\frac{\partial_{t}\gamma}{\gamma}\mp2(D-2)\right)^{-2}\left[\frac{\partial_{y}\gamma}{\gamma}\pm\sqrt{\Big(\frac{\partial_{y}\gamma}{\gamma}\Big)^{2}-4\varepsilon
c^{4}t^{2}\left(t\frac{\partial_{t}\gamma}{\gamma}\mp2(D-2)\right)^{2}}\,\right]\,,
\label{b2} \ee where the $\pm$ now, is independent of the other
two $\mp$'s which, however, go together with those of equations
(\ref{motion}), (\ref{algebraic}). Note that for $\varepsilon=-1$,
the $+$ sign of $\pm$ in (\ref{b2}) corresponds to solutions with
$\partial_{y}\gamma>0$, while the $-$ sign to
$\partial_{y}\gamma<0$.
\par
A remark is necessary here: If we consider the region $y>y(t)$
reversing the normal vector (\ref{normal}), we can check that
expressions (\ref{motion}) and (\ref{algebraic}) remain the same.
\par
If we assume, instead of only $\Lambda_{b}$, a perfect fluid on
the domain wall, expressed by an energy-momentum tensor
$(\rho+p)\mathrm{e}_{A}\mathrm{e}_{B}+ph_{AB}$, then we have to
put on the right hand side of equation (\ref{symmetric}) the
additional term
$-\frac{\kappa_{(D)}^{2}}{2}\left((\rho+p)\mathrm{e}_{A}\mathrm{e}_{B}+\frac{\rho}{D-2}\,h_{AB}\right)$
(with $\varepsilon=-1$). Then, relevant extra matter terms enter
equations (\ref{extrinsic00final}), (\ref{onescale}) through
$K_{00}$, $K$. For a general energy-momentum tensor
$^{(D-1)}T_{AB}$ on the wall, we can, by taking the covariant
derivative with respect to $h_{AB}$ of the Israel equations
(\ref{israel}) and making use of the Codacci's equations and of
the bulk Einstein equations (\ref{einstein}), arrive at the common
conservation law
$h^{A}_{B}h^{C}_{D}\nabla_{C}\,^{(D-1)}T^{D}_{A}=0$. Thus, for
e.g. $p=w\rho$, the modified equations (\ref{onescale}),
(\ref{acceleration}) will contain extra terms for $\gamma_{DW}$.
One has then, to integrate these, finding an expression similar to
(\ref{algebraic}) and proceed further with the bulk equations
checking their compatibility; we are not going to further indulge
into this.

\section*{3 \,\,\,The Solution}
\hspace{0.8cm} In this section we proceed assuming the ansatz
(\ref{ansatz}) for the bulk metric and that the moving domain wall
is a space of \underline{constant curvature} $k\in\{-1,0,1\}$
characterized by a scale factor $a(t,y)$, i.e.
$\gamma=a^{2(D-2)}\gamma_{k}(x)$, where $\gamma_{k}(x)$ is the
determinant of the line element of the constant curvature space.
Since, only $t$ and $y$ logarithmic derivatives of $\gamma$ appear
in the expressions (\ref{algebraic}), (\ref{b2}), $\gamma_{k}(x)$
appears nowhere else. If such a bulk metric exists, admitting the
above described motion of the domain wall, this solution has to be
compatible with equation (\ref{algebraic}). In \cite{vollick}, a
general solution of the Einstein equations with a cosmological
constant in the bulk space for a flat 3-universe has been
obtained, without even assuming the ansatz (\ref{ansatz}).
However, there, the conformal gauge for the $(t,y)$ part of the
metric has been adopted. Since there is no explicit way to go back
to a metric of the form (\ref{linelement}), we cannot exploit this
solution to check the compatibility with the moving domain wall
framework. In \cite{binetruy}, one of the important results
obtained is that the system (for $\varepsilon=-1$, $D=5$) of the
field equations \be R_{AB}-\frac{1}{2}Rg_{AB}=-\Lambda_{B}g_{AB}
 \label{einstein}
  \ee
of our bulk metric, is equivalent to the system of equations
\be
\frac{N\,\acute{}}{N}\,\frac{\dot{a}}{a}+\frac{a\,\acute{}}{a}\,\frac{\dot{b}}{b}-\frac{\dot{a}\,\acute{}}{a}=0
\label{binetruy1} \ee
\be
\frac{\dot{a}^{2}}{N^{2}a^{2}}-\frac{a\,\acute{}\,\,^{2}}{b^{2}a^{2}}=-\frac{k}{a^{2}}+\frac{\mathcal{C}}{a^{4}}+\frac{\Lambda_{B}}{6}\,,
\label{binetruy2}
 \ee
 where $\mathcal{C}$ is a constant of
integration, while the dot and the prime stand for the
$\partial_{t}$ and $\partial_{y}$ derivatives respectively.
Equations (\ref{binetruy1}), (\ref{binetruy2}) for the ansatz
(\ref{ansatz}) and the time choice (\ref{gauge}) get the following
form
\be
\frac{a\,\acute{}}{a}\,\frac{\dot{b}}{b}-\frac{\dot{a}\,\acute{}}{a}=0
\label{bin1}
 \ee
 \be
b^{2}\dot{a}^{2}-\frac{a\,\acute{}\,\,^{2}}{b^{2}}=F, \label{bin2}
\ee where
\be
F=-k+\frac{\mathcal{C}}{a^{2}}+\frac{\Lambda_{B}}{6}\,a^{2}.
\label{F}
 \ee
These equations together with (\ref{algebraic}) is everything we
have to satisfy. Equation (\ref{bin1}) is easily integrated to
\be
a\,\acute{}=b(t)A(y),
 \label{binet1}
  \ee
where $A(y)$ is an arbitrary function of $y$. Then, equation
(\ref{bin2}) becomes
\be
\dot{a}^{2}=\frac{A^{2}+F}{b^{2}}\,.
 \label{binet2}
  \ee
Substituting $a\,\acute{}\,,\dot{a}$ from (\ref{binet1}),
(\ref{binet2}) into (\ref{algebraic}), we obtain equivalently the
following algebraic equation:
\be
c^{2}t^{2}F-(c^{2}a^{2}+A^{2})b^{2}\pm 2 c A a b
\sqrt{b^{2}+c^{2}t^{2}}=0.
 \label{algebraicnew}
 \ee
 (The signs $\pm$ appearing in this equation do not necessarily correspond to
 those of equations (\ref{motion})-(\ref{b2})). Note again, that the system has
been reduced to equations (\ref{binet1}), (\ref{binet2}) and
(\ref{algebraicnew}).
\par
In order to proceed further, we will restrict ourselves to the
case \underline{$\mathcal{C}=0$} (for fixed-brane cosmologies with
cosmological constant in the bulk and even perfect fluid on the
brane, this was shown \cite{mohapatra} to be the general case) and
\underline{$k=0$} (flat 3-universe). In this case, we can solve
equation (\ref{algebraicnew}) getting
\be
a=\frac{\sqrt{b^{2}+c^{2}t^{2}}\pm
\sqrt{c^{2}+\frac{\Lambda_{B}}{6}}\,\,t}{c\,(b^{2}-\frac{\Lambda_{B}}{6}t^{2})}\,bA.
\label{scale}
\ee
These solutions arise from the $+$ sign of
equation (\ref{algebraicnew}). The other choice of sign in that
equation makes the solution (\ref{scale}) to change the overall
sign.
\newline
It is obvious from (\ref{scale}) that the only way the model can
possess a solution (if this really exists) is of a separable form
for $a(t,y)$. (If we consider $k\neq0$ this does not happen any
more). Differentiating (\ref{scale}) with respect to $y$ and
substituting in (\ref{binet1}) we obtain
\be
\frac{A\,\acute{}}{A}=\frac{c\,(b^{2}-\frac{\Lambda_{B}}{6}t^{2})}{\sqrt{b^{2}+
c^{2}t^{2}}\pm\sqrt{c^{2}+\frac{\Lambda_{B}}{6}}\,\,t}\,\,.
\label{separation} \ee
 Thus, the only way this equation can be
satisfied is when each side is a non-zero constant, say
$\theta\neq0$. For the $-$ sign of equation (\ref{algebraicnew}),
this is still correct for the quantity appearing on the right-hand
side of (\ref{separation}), but the quantity on the left-hand side
is equal to $-\theta$.
\newline
From (\ref{scale}) we get \be a(t,y)=\frac{1}{\theta}\,b(t)A(y).
 \label{a}
  \ee
The solution for $A(y)$ coming from (\ref{separation}) is
\be
A(y)=A_{o}\,e^{\theta y},
 \label{A}
  \ee
with $A_{o}$ being a constant of integration.
\newline
Since we are basically concerned with $a^{2}$ and not $a$, we
disregard the $-$ sign arising from (\ref{algebraicnew}), but we
keep in mind that the only alteration this causes to the solutions
is the change of $\theta$ to $-\theta$ in the $y$-exponent of $A$
(or even $a$).
\newline
Differentiating equation (\ref{a}) with respect to $t$ and
substituting in (\ref{binet2}) we obtain \be
\dot{b}^{2}=\frac{\Lambda_{B}}{6}+\frac{\theta^{2}}{b^{2}}\,.
\label{bdot2} \ee The solutions for $b(t)$ coming from the
right-hand side of (\ref{separation}) and agreeing with
(\ref{bdot2}) are \be
b^{2}(t)=\frac{\Lambda_{B}}{6}\,t^{2}\pm\frac{2
\theta}{c}\,\sqrt{c^{2}+\frac{\Lambda_{B}}{6}}\,\,t+\frac{\theta^{2}}{c^{2}}\,.
\label{bsquare} \ee Gathering together all the different cases and
rescaling $a^{2}$, we can write the final solution of our system
as follows \be b^{2}(t)=\frac{\Lambda_{B}}{6}\,t^{2}+\mu
\,\sqrt{\Lambda_{b}^{2}+6\Lambda_{B}}\,\,t+9\mu^{2}.
 \label{b2f}
 \ee
\be
a^{2}(t,y)=\left(\frac{\Lambda_{B}}{6}\,t^{2}+\mu
\,\sqrt{\Lambda_{b}^{2}+6\Lambda_{B}}\,\,t+9\mu^{2}\right)e^{\pm\mu
\Lambda_{b}y}\,,
 \label{a2}
  \ee
containing the 3 parameters $\mu$, $\Lambda_{b}$ and $\Lambda_{B}$
(we have set $\mu=\frac{2\theta}{\Lambda_{b}}\neq0$). Certainly,
the solution exists for $\Lambda_{b}^{2}+6\Lambda_{B}>0$. Note
that, because of the procedure followed, the above solution is the
unique one for $\mathcal{C}=k=0$ and $N=N(t)$, $b=b(t)$.
\par
 On spacetime sections $dy=0$, our bulk solution obviously reduces to a
spacetime conformal to Minkowski space; however, it is not locally
$dS_{4}$ or $AdS_{4}$. The bulk solution obtained by (\ref{b2f}),
(\ref{a2}) can be seen to have vanishing Weyl tensor, hence it is
conformally flat. Furthermore, since equations (\ref{einstein})
hold, the bulk space is a space of constant curvature, thus, it is
locally $dS_{5}$ or $AdS_{5}$. So, expressions (\ref{b2f}) and
(\ref{a2}) define an embedding of a conformally Euclidean
four-dimensional space as codimension one hypersurfaces of such a
higher-dimensional space.
 \par
 It may be of some importance to note that in
\cite{binetruy} (or \cite{mohapatra}), assuming a different ansatz
for the metric (\ref{linelement}) (i.e. $b=b(y)$ and thus choosing
$b=1$), a separable form for the scale factor $a$ with truly
exponential $y$-dependence can be obtained only through the fine
tuning between $\Lambda_{b}$ and $\Lambda_{B}$ appearing in
\cite{randall} and only for $\Lambda_{B}<0$. A similar exponential
damping appears in our case without any fine tuning and even for
$\Lambda_{B}>0$; however, it does not play the role of a warp
factor of a four-dimensional Minkowski (\cite{randall}) or de
Sitter (\cite{dewolfe}) space. In the same context, we note the
following: If $\zeta$ is an affine parameter for the transversal
geodesics emanating from the domain wall, we see from equation
(\ref{time}) below (making use of (\ref{normal}), (\ref{b2f}),
(\ref{y})), that, apart from the overall factor $b^{2}(t)$, the
``inner'' exponential factor $e^{\pm \mu \Lambda_{b} y}$ is
proportional to $\exp(\mp
\frac{\mu\Lambda_{b}}{6b^{2}}\sqrt{36b^{2}+\Lambda_{b}^{2}t^{2}}\,\zeta)$,
and decreases, for suitable sign of $\mu\Lambda_{b}$, with respect
to $\zeta$ as we move away from the wall.
 \par
 In the context of non-compactified Kaluza-Klein
theory, a solution of the form (\ref{time}) has been given in
\cite{leon}, where a different time-warp factor appears, since
that refers to zero $\Lambda_{B}$.

\section*{4 \,\,\,Three-Universe Evolution}
\hspace{0.8cm} The bulk metric found in the preceding section
allows us to investigate the trajectory of the moving domain wall
in the bulk space as well as its own time evolution. The evolution
of the three-universe is determined by the time evolution of the
moving domain wall scale factor, denoted by $a_{DW}$ (see
(\ref{induced metric})). The time-dependence of $a_{DW}$ enters
firstly, through the explicit time-dependence of the bulk scale
factor $a$ and secondly, due to the $y(t)$ dependence of $a$ as we
follow the trajectory of the domain wall through the bulk space,
i.e. \be a_{DW}^{2}(t)=a^{2}(t,y(t)).
 \label{adw}
 \ee
The function $y(t)$ is found, after integrating equation
(\ref{motion}) with the help of (\ref{b2f}), to be
\be
y(t)-y_{o}=\pm\frac{1}{\mu
\Lambda_{b}}\ln\frac{(\sqrt{\Lambda_{b}^{2}+6\Lambda_{B}}\,\,t+18\mu)^{2}}{b^{2}(t)}\,,
\label{y}
 \ee
with $y_{o}$ a constant of integration. Because of the
positiveness of $b^{2}$, we conclude from (\ref{b2f}) that the
range $\mathcal{D}$ of definition of time $t$ is
$(-\infty,t_{1})\cup(t_{2},+\infty)$ for $\Lambda_{B}>0$ and
$(t_{1},t_{2})$ for $\Lambda_{B}<0$, where $t_{1}$, $t_{2}$ are
the two distinct roots of the expression (\ref{b2f}) for
$b^{2}(t)$. If one attempted to go beyond the above time
intervals, then $b^{2}$, $a^{2}$ would become negative and thus,
the bulk would seem to have four timelike directions (a similar
situation in common T-NUT-M space \cite{ryan} or even
Schwarzschild spacetime interchanges the role played by time with
one spacelike coordinate).
\newline
For $\Lambda_{B}<0$, the domain wall motion (\ref{y}) is obviously
bounded in the $(t,y)$ plane. For $\Lambda_{B}>0$,
$y(t)\rightarrow$ constant as $t$ goes to infinity, and the motion
develops inside a strip in the $(t,y)$ plane. This asymptotic
behavior resembles quantitatively the behavior of horospheres or
equidistant hypersurfaces (the unique totally umbilic
hypersurfaces excluding totally geodesic ones that are
hyperplanes, and geodesic spheres being bounded) of hyperbolic
spaces (see \cite{spivak}). Making a coordinate transformation
mixing $t,y$ in the $(A)dS_{5}$ space, all the found expressions
will change form; nevertheless, the geometrical domain wall object
is unaffected and the same happens for the range of values of
$a_{DW}$.
\par
Substitution of expression (\ref{y}) in equation (\ref{adw}) gives
for $a_{DW}$:
\be
a_{DW}^{2}(t)=e^{\pm\mu
\Lambda_{b}y_{o}}(\sqrt{\Lambda_{b}^{2}+6\Lambda_{B}}\,\,t+18\mu)^{2}.
\label{adow}
 \ee
 The double sign in the exponent corresponds to the double sign of
 (\ref{a2}). It can be easily checked that the value $t^{\ast}$ of time, for which expression
 (\ref{adow}) vanishes, does not belong in $\mathcal{D}\cup\{t_{1},t_{2}\}$ for any
non-zero $\Lambda_{B}$. Thus, $a_{DW}$ never vanishes (or even
tends to zero), which means that this three-universe will never
collapse to a singularity. The qualitative behavior of $a_{DW}$ is
shown, for the various cases of the parameters $\Lambda_{B}$,
$\mu$, in the diagrams of figure 1.
%
%
\begin{figure}[h!]
\centering
\includegraphics*[width=149pt, height=100pt]{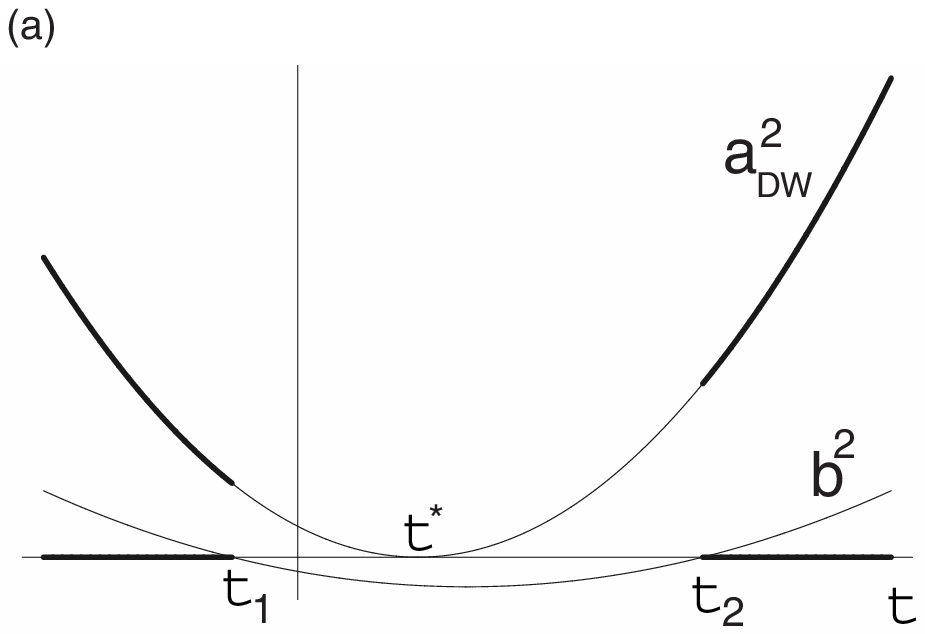}%
\hspace{0.1cm}%
\includegraphics*[width=149pt, height=100pt]{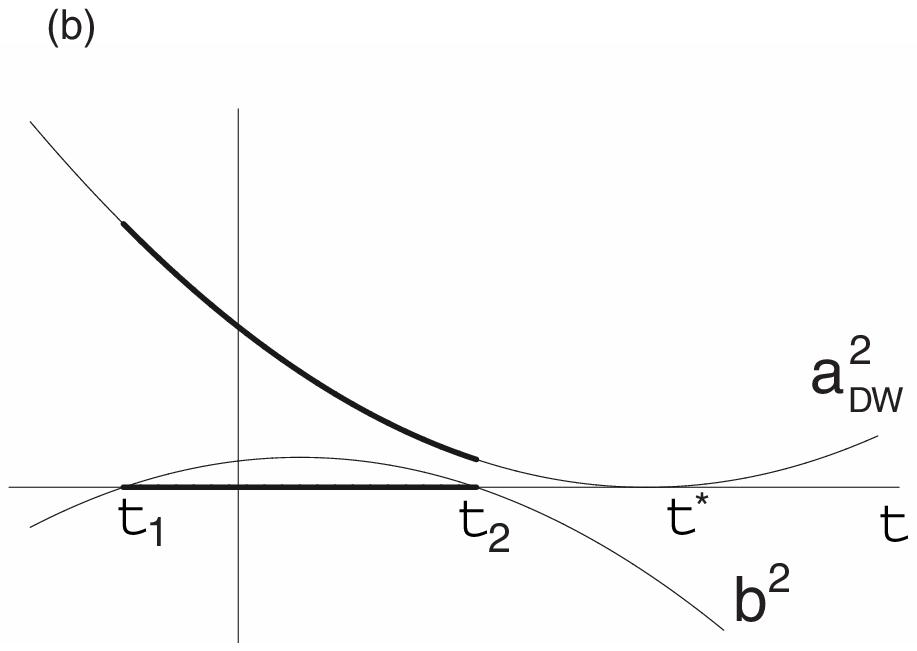}%
\hspace{0.1cm}%
\includegraphics*[width=149pt, height=100pt]{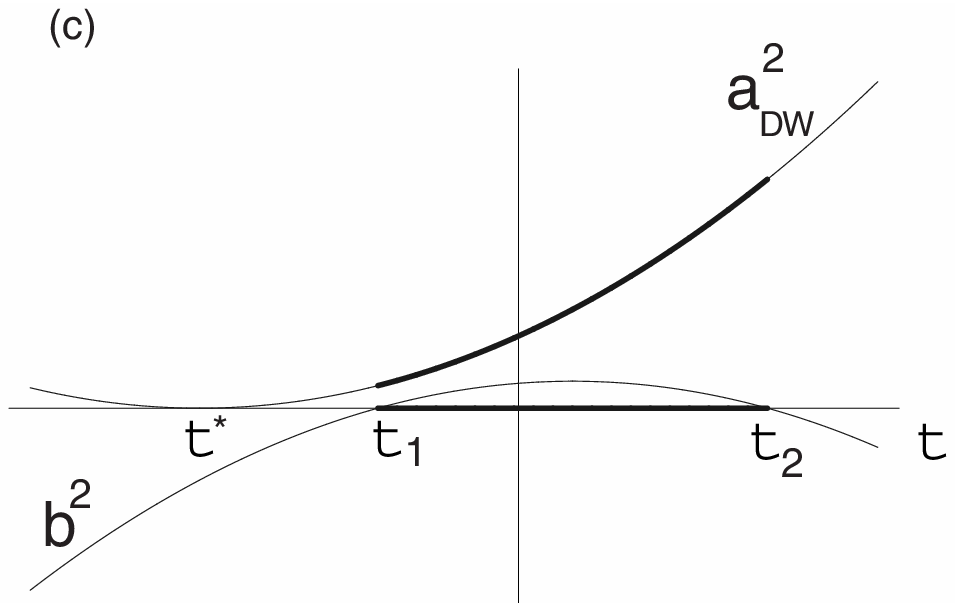}
\caption{(a) $\Lambda_{B}>0$\,\,\,\,\,,\,\,\,\,\,(b)
$\Lambda_{B}<0$ and $\mu<0$\,\,\,\,\,,\,\,\,\,\, (c)
$\Lambda_{B}<0$ and $\mu>0$}
\end{figure}
\par
 Since in the solution (\ref{b2f})-(\ref{a2}) (as it is also seen from equation (\ref{a})) the
time-dependent part of $a^{2}$ is exactly $b^{^2}$, there exists a
time parameter $\eta$ (``conformal time''), such that the whole
time-dependence $b^{2}(\eta)\equiv b^{2}(t(\eta))$ of the
line-element (\ref{linelement}) is factorizable, i.e.
\be
ds^{2}=b^{2}(\eta)(-d\eta^{2}+dy^{2}+e^{\pm \mu\Lambda_{b}y
}\delta_{ij}dx^{i}dx^{j}).
 \label{time}
  \ee
As $t$ approaches the endpoints $t_{1}, t_{2}$, the scale factor
$b^{2}$ tends to zero. How can a non-zero scale factor of the
three-universe exist in this limiting region? This vanishing is
characteristic of the foliation of the bulk space defined by the
above line-element; the bulk space, as explained in the previous
section, is for all times a space of constant curvature. However,
the moving three-universe resides in a hypersurface which does not
belong to this foliation. So, what we have found is that
$a_{DW}(t)\rightarrow$ constant as $t\rightarrow t_{1},t_{2}$,
even though $a(t,y)\rightarrow 0$ for any fixed $y$. Further use
of the form (\ref{time}) for the metric will be made in section 5.
\par
As far as proper time $\tau$ is concerned, we get from equations
(\ref{time-proper time}), (\ref{motion}) and (\ref{adow}) that
\be
a_{DW}^{2}(\tau)=e^{\pm\frac{1}{3}\sqrt{\Lambda_{b}^{2}+6\Lambda_{B}}\,(\tau-\tau_{o})}\,.
\label{tau}
 \ee
 The proper time follows $t$ to $\pm\infty$ but
remains finite as $t\rightarrow t_{1},t_{2}$ for the case
$\Lambda_{B}>0$, while it is still defined in a finite interval
for $\Lambda_{B}<0$. Note that the $+$ sign of the preceding
expression refers to the cases (i) $\Lambda_{B}<0$ and $\mu>0$,
(ii) $\Lambda_{B}>0$ and for the $+\infty$ region solution, while
the $-$ sign refers to (iii) $\Lambda_{B}<0$ and $\mu<0$, (iv)
$\Lambda_{B}>0$ and for the $-\infty$ region solution. Two more
remarks concerning equation (\ref{tau}) are pertinent: (1) This
three-universe solution has a ``classical'' four-dimensional de
Sitter form with an ``effective'' cosmological constant
$\Lambda_{dS}=\frac{1}{12}(\Lambda_{b}^{2}+6\Lambda_{B})$. (2)
This solution, in contrast to the conventional de Sitter one, does
not collapse (see e.g. \cite{hawking}), a fact that certainly
comes from the existence of extra dimensions. Thus, the domain
wall with the metric (\ref{induced metric}) is a portion of
$dS_{4}$ space (embedded in $(A)dS_{5}$). The full domain wall
extension is the geodesically complete and non-singular $dS_{4}$
space.
\par
In fixed-brane cosmologies, there is an autonomous equation
\cite{binetruy} governing the time evolution of the scale factor
of the three-universe, which is obtained without having knowledge
of the metric outside the brane. On the contrary, in the moving
formalism, the exact solution of the bulk space is necessary in
order to specify the three-universe evolution, a fact due to the
dynamical interaction of the domain wall with the entire space.
Furthermore, for a prefixed-position brane, an autonomous induced
dynamics on the brane was obtained \cite{maeda} for an $(A)dS_{5}$
or Minkowski bulk. Although this fact does not concern our
procedure, after we have resulted in an $(A)dS_{5}$ bulk, both
conclusions must be in agreement. Actually, the matter content of
the brane supplied by \cite{maeda} is exactly our ``effective''
cosmological constant.

\section*{5 \,\,\,Scalar field description}
\hspace{0.8cm} It would be interesting for a further investigation
of the physical properties of the bulk metric to consider some
tensor perturbations of it. We do not embark on this study here,
but instead we deal, in a simplified setting, with the inclusion
of a scalar field propagating (but not back-reacting) in the
spacetime described by equations (\ref{b2f}), (\ref{a2}). Although
the bulk background can be cast into a standard $(A)dS_{5}$ form,
we believe that the consideration of such a field through the
foliation arising from the privileged role played by the
three-universe is meaningful. As a starting comment, we recall
that it was proved in \cite{dewolfe} that the equations governing
the transverse traceless fluctuations around the general classical
five-dimensional background metric with four-dimensional
Poincar$\acute{e}$ symmetry (which is not certainly the situation
we intent to investigate) coincide with the equation of motion of
a free massless scalar field in the same curved background. In an
$AdS_{5}$ black hole background, the spin-2 components of the
graviton were also found to obey a free scalar wave equation
\cite{constable,brower}.
\newline
The action of our bulk field $\phi(t,y,x^{i})$ is assumed to be
\be
S=\frac{1}{2}\int_{M} d^{5}x \sqrt{-g}\,
(g^{AB}\partial_{A}\phi\partial_{B}\phi+m^{2}\phi^{2})+\int_{\Sigma}
d^{4}x \sqrt{-h}\,\hat{V}(\phi),
 \label{action}
  \ee
where $g=get(g_{AB})$, $h=det(h_{AB})$, $M$ is the full
five-dimensional spacetime and $\Sigma$ is the domain wall
hypersurface. Setting the variation of the action with respect to
$\phi$ equal to zero yields the scalar field equation
\be
(\Box-m^{2})\phi=\frac{d\hat{V}}{d\phi}\delta(x^{A}-x^{A}_{DW}),
 \label{box}
  \ee
where
$\Box=\frac{1}{\sqrt{-g}}\partial_{A}(\sqrt{-g}\,g^{AB}\partial_{B})$
is the five-dimensional Laplacian. The presence of the delta
function in equation (\ref{box}) is irrelevant for the integration
procedure outside $\Sigma$, but it has to be included at the
boundary condition on the hypersurface, if one wants to get a
global solution. In the same spirit, Israel junction conditions
are the boundary conditions for the geometry. The role played by
the potential $\hat{V}(\phi)$ appearing in (\ref{action}) will be
explained later. The importance of such surface terms, added to
bulk scalar field actions without back-reaction, has been
discussed, within the context of fixed-brane cosmologies, in
\cite{goldberger} (the back-reaction has been taken into account
in \cite{dewolfe, tetradis}). Integrating equation (\ref{box})
around $\Sigma$, we get the following boundary condition for the
continuous field $\phi$: \be
[n^{A}\partial_{A}\phi]=\left(\frac{d\hat{V}}{d\phi}\right)_{wall}\,,
 \label{boundary}
\ee where the bracket denotes, as usual, the discontinuity of the
quantity across the domain wall.
\par
We will work out equation (\ref{box}) using as time-parameter the
conformal time $\eta$ of the bulk metric defined in (\ref{time}).
Then, the wave equation can be written as
\be
\left(\frac{\partial_{\eta}^{2}\phi}{\phi}+\partial_{\eta}(\ln
b^{3})\,\frac{\partial_{\eta}\phi}{\phi}+m^{2} b^{2}\right)
-\left(\frac{\phi\,\acute{}\,\acute{}}{\phi}\pm
\frac{3}{2}\mu\Lambda_{b}\,\frac{\phi\,\acute{}}{\phi}\right)-e^{\mp\mu\Lambda_{b}y}\,\frac{\delta^{ij}\partial_{i}\partial_{j}\phi}{\phi}=0.
\label{wave}
 \ee
Because of the homogeneity of the spatial sections, the
$x$-dependence of the mode solutions are separable, as it is seen
from (\ref{wave}), i.e.
\be
\phi_{\vec{k}}(\eta,y,x^{i})=(2\pi)^{-\frac{3}{2}}\,e^{i\vec{k}\cdot\vec{x}}\,\tilde{\phi}_{k}(\eta,y),
 \label{xsector}
 \ee
where $k=(\delta_{ij}k^{i}k^{j})^{\frac{1}{2}}$ and the prefactor
in (\ref{xsector}) guarantees the standard normalization of its
spatial part. Then, the field equation (\ref{wave}) admits
separability in $\eta$ and $y$ as well, i.e.
\be
\tilde{\phi}_{k}(\eta,y)=u_{k}(\eta)\psi_{k}(y)
 \label{tysector},
\ee where $\psi_{k}$, $u_{k}$ satisfy respectively the equations
\be
\frac{\psi_{k}\acute{}\,\acute{}}{\psi_{k}}\pm\frac{3}{2}\mu\Lambda_{b}\,\frac{\psi_{k}\acute{}}{\psi_{k}}-k^{2}\,e^{\mp\mu\Lambda_{b}y}+
\lambda=0 \label{psi} \ee
\be
\frac{\partial_{\eta}^{2}u_{k}}{u_{k}}+\partial_{\eta}(\ln
b^{3})\,\frac{\partial_{\eta}u_{k}}{u_{k}}+m^{2} b^{2}+ \lambda=0,
 \label{u}
\ee
 with $\lambda$ being a new separation constant. The first
of these equations is transformed through a combined change of
$y$, $\psi_{k}$\,:
 \be
\bar{y}=\frac{2k}{|\mu\Lambda_{b}|}\,e^{\mp\frac{1}{2}\mu\Lambda_{b}y}\,\,\,,\,\,\,
\bar{\psi}_{k}=e^{\pm\frac{3}{4}\mu\Lambda_{b}y}\,\psi_{k},
\label{psipsibar}
 \ee to the modified Bessel equation
\be
\bar{y}^{2}\,\frac{d^{2}\bar{\psi}_{k}}{d\bar{y}^{2}}+\bar{y}\,\frac{d\bar{\psi}_{k}}{d\bar{y}}-(\bar{y}^{2}+\frac{9}{4}-\frac{4\lambda}{\mu^{2}\Lambda_{b}^{2}})\,\bar{\psi}_{k}=0.
\label{bessel} \ee
 The solutions to (\ref{bessel}) consist of
linear combinations of the modified Bessel functions
$I_{\rho}(\bar{y})$, $K_{\rho}(\bar{y})$, where
$\rho=\sqrt{\frac{9}{4}-\frac{4\lambda}{\mu^{2}\Lambda_{b}^{2}}}$\,\,.
\par
As a consequence of the signature of the bulk metric $g_{AB}$, the
wave operator is hyperbolic. Thus, we do not expect (and even
want), square integrability of $\phi$ with respect to time under
the hermitian measure of (\ref{wave}). Nevertheless, integrability
with respect to $x^{i},\,y$, i.e. on the constant time
hypersurfaces, would be desirable. Due to the fact that the
classical domain wall motion in the bulk space, as described by
the function $y(t)$ of (\ref{y}), is bounded, $\psi_{k}(y)$ will
be normalizable in this compact $y$-range $\mathrm{D}$, with
either the flat measure or the natural curved background one
$e^{\pm \frac{3}{2} \mu \Lambda_{b} y}$. However, this is not the
final word. In the treatments with $y=$constant branes, the
three-universe is unfolded in one such hypersurface; no motion
takes place, precisely speaking, in the extra dimension, whose
existence certainly influences the 3-motion. On the contrary, in
our model the true classical motion of the three-universe takes
place in an unbounded five-dimensional space. Moreover, having in
mind that the classical $\phi$ will have a quantum mechanical
analogue, we state that it is more desirable for $\psi_{k}$ to be
normalizable in the infinite $y$-range. Of course, the $t$-time
range of the bulk solution is $\mathcal{D}$. The functions
$I_{\rho}$\,, $K_{\rho}$ or any linear combination of them have an
exponentially damping behaviour in the neighborhood of one of the
two infinities, but not both. The existence of the domain wall can
guarantee this normalizability. The boundary condition
(\ref{boundary}) gives information only for the discontinuity of
the normal to the wall component of the gradient of $\phi$. Its
parallel components can be either continuous or discontinuous. The
existence of $\hat{V}(\phi)$ in (\ref{action}) and the
non-determination of $[\mathrm{e}^{A}\,\partial_{A}\phi]$ from the
above condition, allows us to obtain $[\partial_{\eta}u_{k}]=0$
for any $\eta$ and $[\partial_{y}\psi_{k}]\neq 0$ for few or any
values of $y\in \mathrm{D}$. As it can be seen from
(\ref{boundary}), the necessary and sufficient condition for
$[\partial_{\eta}u_{k}]=0$ is
\be
[\mathrm{e}^{A}\,\partial_{A}\phi]=\frac{\mathrm{e}_{y}}{n_{y}}\left(\frac{d\hat{V}}{d\phi}\right)_{wall}.
\label{tangentboundary}
 \ee
Thus, $u_{k}(\eta)$ is differentiable at any instant of time and
equation (\ref{u}) can be investigated further. Then, if
$(d\hat{V}/d\phi)_{wall}$ is not identically zero, there exists at
least one of the $\psi_{k}$'s with $[\partial_{y}\psi_{k}]\neq 0$,
at least at one point $y$. In this way, $\psi_{k}(y)$ is managing
to ``turn'' its slope down and assure normalizability in the
infinite $y$-range. More sophisticated situations could arise
depending on the form of $(d\hat{V}/d\phi)_{wall}$. It is possible
even to obtain functions $\psi_{k}$ continuous in $\mathrm{D}$,
nowhere differentiable. The restriction of $\phi$ on the wall
(which could be interpreted as the restriction of $\phi$ on our
visible universe) will be a continuous function of $t,x^{i}$,
possessing non-differentiable in time characteristics due to its
part $\psi_{k}(y(t))$.
\par
So much about $\psi_{k}$. Now consider equation (\ref{u}).
Defining the function
\be
 \bar{u}_{k}=b^{\frac{3}{2}}\,u_{k},
 \label{canonical}\ee
we transform this equation to its canonical form
\be
\frac{\partial_{\eta}^{2}\bar{u}_{k}}{\bar{u}_{k}}+\left[\left(m^{2}-\frac{5\Lambda_{B}}{8}\right)
b^{2} +\left(\lambda-\frac{9
\mu^{2}\Lambda_{b}^{2}}{16}\right)\right]=0,
 \label{ubar} \ee
in which use of equation (\ref{bdot2}) has been made. We can find
from (\ref{bdot2}) the exact relation between $b$ and $\eta$ to be
\be
b(\eta)=\frac{|\mu \Lambda_{b}|}{2} \sqrt{\frac{6}{\Lambda_{B}}}\,
\csc h (\mp \frac{| \mu \Lambda_{b}|}{2} \,\eta),
 \label{b-eta}
 \ee
 \be
b(\eta)=\frac{|\mu \Lambda_{b}|}{2}
\sqrt{\frac{6}{|\Lambda_{B}|}}\, \sec h (\frac{| \mu
\Lambda_{b}|}{2} \,\eta)
 \label{b-etan}
 \ee
 for $\Lambda_{B}>0$ and $\Lambda_{B}<0$ respectively.
 The $-$ ($+$) sign in (\ref{b-eta}), with $\eta$-range $(-\infty,0)$ (resp. $(0,+\infty)$), corresponds to the $+\infty$ ($-\infty$)
 region of the solution (\ref{b2f})-(\ref{a2}), while for $\Lambda_{B}<0$ we have $\eta\in(-\infty,+\infty)$.
 The above expressions, substituted in equation (\ref{ubar}), supply the exact equations for $\bar{u}_{k}$ in $\eta$-time.
Making the transformation
\be
z=\cosh(\frac{|\mu\Lambda_{b}|}{2}\,\eta)\,\,,\,\,F_{k}=\sinh^{-\frac{1}{2}}(\frac{|\mu\Lambda_{b}|}{2}\,\eta)\,\,\bar{u}_{k}\,,
 \label{zeta}
 \ee
 for $\Lambda_{B}>0$, and
\be
z=\cosh(\frac{|\mu\Lambda_{b}|}{2}\,\eta+\frac{i\pi}{2})\,\,,\,\,F_{k}=\cosh^{-\frac{1}{2}}(\frac{|\mu\Lambda_{b}|}{2}\,\eta)\,\,\bar{u}_{k}\,,
\label{F}
\ee
 for $\Lambda_{B}<0$, equation (\ref{ubar}) is converted in both cases to the associated Legendre equation
\be
(1-z^{2})\frac{d^{2}F_{k}}{dz^{2}}-2z\frac{dF_{k}}{dz}+\left[\nu(\nu+1)-\frac{\mathrm{m}^{2}}{1-z^{2}}\right]F_{k}=0,
\label{legen} \ee
 where
$\mathrm{m}=(4-\frac{6m^{2}}{\Lambda_{B}})^{\frac{1}{2}}$\,, and
$\nu(\nu
+1)=2-\frac{4\lambda}{\mu^{2}\Lambda_{b}^{2}}$\,\,$(\mathrm{m},\nu
\in\mathbf{C})$. It is well known that the general solution of
equation (\ref{legen}) consists of linear combinations of Legendre
functions of first and second kind, i.e. $P_{\nu}^{\mathrm{m}}(z)$
and $Q_{\nu}^{\mathrm{m}}(z)$ (in our case toroidal functions). Of
course, the specific choice of this combination determines the
bulk vacuum. Each of these functions carries the real continuous
``separation'' label $\lambda$, contained in $\nu$, which makes
the corresponding $\bar{u}_{k\lambda}$'s (modes) - along with
their complex conjugates - complete sets in the space of
functions. For example, the state $|\,0\,\rangle_{P}$ defined with
respect to the $P$ functions will be inequivalent to the state
$|\,0\,\rangle_{Q}$ defined with respect to the modes $Q$.
\par
When $b^{2}\rightarrow0$, equation (\ref{ubar}) is approximated by
\be
\frac{\partial_{\eta}^{2}\bar{u}_{k}}{\bar{u}_{k}}+\left(\lambda-\frac{9
\mu^{2}\Lambda_{b}^{2}}{16}\right)=0 ,
 \label{approx}
 \ee
which admits for $\lambda > \frac{9\mu^{2}\Lambda_{b}^{2}}{16}$
the ``Minkowski-type'' modes
\be
\bar{u}_{k}^{(\infty)}(\eta)=\frac{1}{\sqrt{2
\omega}}\,e^{-i\omega\eta}\,,
 \label{minkowski}
  \ee
where
$\omega=\sqrt{\lambda-\frac{9\mu^{2}\Lambda_{b}^{2}}{16}}$\,\,.
For $\Lambda_{B}<0$, the regions $b^{2}\rightarrow 0$ are the two
asymptotic regions of (\ref{b-etan}) as
$\eta\rightarrow\pm\infty$. The modes $\bar{u}_{k}$ that behave
like positive frequency modes (\ref{minkowski}) in these limits,
i.e. $\bar{u}_{k}(\eta)\rightarrow \bar{u}_{k}^{(\infty)}(\eta)$
as $\eta\rightarrow \pm \infty$, are
\be
\bar{u}_{k}(\eta)=\frac{(-1)^{-\frac{1}{2}\mathrm{m}}\,i^{\nu+\mathrm{m}+1}}{\sqrt{\pi\omega}\,
e^{i\mathrm{m}\pi}}\frac{\Gamma(\nu+\frac{3}{2})}{\Gamma(\nu+\mathrm{m}+1)}\cosh^{\frac{1}{2}}
(\frac{|\mu\Lambda_{b}|}{2}\eta)\,Q_{\nu}^{\mathrm{m}}(i\sinh(\frac{|\mu\Lambda_{b}|}{2}\eta)).
\label{lambdanegative}
 \ee
 This expression holds for $\eta>0$ (``out'' modes),
while for $\eta<0$ (``in'' modes) it has to be multiplied by
$(-1)^{\nu-\mathrm{m}}$. For $\eta>0$, Im$\nu>0$ is used, while
for $\eta<0$, Im$\nu<0$. Vacuum states defined in similar ways
have been given in \cite{rumpf,dowker}. For the derivation of
equation (\ref{lambdanegative}) the defining relation of
\,$Q_{\nu}^{\mathrm{m}}(z)$\, in terms of\,
$_{2}F_{1}\,(1+\frac{1}{2}\,(\nu+ \mathrm{m}\,)\,,\,
\frac{1}{2}\,(1+\nu+
\mathrm{m}\,)\,;\,\nu+\frac{3}{2}\,;\,\frac{1}{z^{2}})$\,,\,
$|z|>1$ has been used (see \cite{abramo}, p. 332). The modes
$\bar{u}_{k}^{in}$, $\bar{u}_{k}^{out}$ are connected through a
Bogolubov transformation, thus, if the quantum state is chosen to
be $|\,0\,\rangle_{in}$, an unaccelerated particle detector in the
out region will detect some spectrum. Of course, the detector is
considered to move in the bulk space and the particle creation
occurs in this space due to the bulk cosmological evolution.
Further examination of the situation would require  the evaluation
of the Wightman function in the out region constructed using the
in vacuum.
\par
It holds that $|\partial_{\eta}(\ln b^{2})|\rightarrow
|\mu\Lambda_{b}|$ as $\eta\rightarrow\infty$ and
$\partial_{\eta}(\ln b^{2})\rightarrow \infty$ as $\eta\rightarrow
0$. Thus, since our spacetime is not slowly expanding in the two
asymptotic limits, we cannot define physically reasonable
adiabatic in and out vacua. Although adiabatic in and out regions
do not exist, it is still possible to define adiabatic vacua as
being those which are vacuous in the high label $\lambda$ modes.
Recall \cite{birrell} that for the case of our harmonic
oscillator-type equation (\ref{ubar}) (with time-dependent
frequency), the ``zeroth order adiabatic modes''
\be
\bar{u}_{k}^{(0)}(\eta)=\frac{1}{\sqrt{2}\,\,\Omega_{\lambda}^{1/4}(\eta)}\,e^{-i\int
^{\eta}\Omega_{\lambda}^{1/2}(\eta')d\eta'},
 \label{adiabatic}
\ee where
$\Omega_{\lambda}(\eta)=(\lambda-\frac{9\mu^{2}\Lambda_{b}^{2}}{16})+(m^{2}-\frac{5\Lambda_{B}}{8})b^{2}(\eta)$,
become good approximations to exact adiabatic positive frequency
modes, when the quantity
\be
T^{2}\Omega_{\lambda}(\eta_{1})=T^{2}\left(\lambda-\frac{9\mu^{2}\Lambda_{b}^{2}}{16}\right)\pm
T^{2}\,\frac{3\mu^{2}\Lambda_{b}^{2}}{2}\left(\frac{m^{2}}{\Lambda_{B}}-\frac{5}{8}\right)\csc
h ^{2}\left(\frac{|\mu\Lambda_{b}|}{2}\,\eta_{1}T\right)
\label{goodapprox}
 \ee
 - with $T$ being the so-called adiabatic
parameter and $\eta_{1}=\frac{\eta}{T}$ - becomes large with
respect to the derivatives of $\partial_{\eta}\ln
b^{2}=$O$(T^{-1})$ for fixed $\eta_{1}$. This is the case for
large $\lambda$ or large $m^{2}/|\Lambda_{B}|$ (but not for small
$|\eta|$), either individually or together. In (\ref{goodapprox})
the $+\,(-)$ sign corresponds to $\Lambda_{B}>0\,(<0)$ and the
$\csc h$ has to be replaced by $\sec h$ in the $\Lambda_{B}<0$
case. The above described picture resembles somehow the common
4-dimensional quantum theory of de Sitter space, where the
trigonometric functions have been replaced by hyperbolic ones. In
the limit of large $\lambda$ with $\mathrm{m}$, $\mu\Lambda_{b}$
and $\eta$ fixed, we have \,$\Omega_{\lambda}(\eta)\simeq
\lambda$, which, when substituted in (\ref{adiabatic}), gives
\be
\bar{u}_{k}^{(0)}(\eta)\rightarrow\frac{1}{\sqrt{2}\,\lambda^{\frac{1}{4}}}\,e^{-i\sqrt{\lambda}\,\eta}\,.
\label{asympto}
 \ee
\par
For $\Lambda_{B}>0$, the choice
 \bea
 \bar{u}_{k}(\eta)=&\frac{(\nu +\frac{1}{2})^{\frac{1}{2}- \mathrm{m}}}{2^{2
\mathrm{m}+1}\,\lambda^{\frac{1}{4}}}\,\frac{\Gamma(1-
\mathrm{m})\,\Gamma(\frac{1}{2}- \mathrm{m})}{f\,\Gamma(1-2
\mathrm{m})} \,\sinh^{\frac{1}{2}}(\frac{| \mu \Lambda_{b}|}{2}
\eta)\,[\frac{f-1}{\sin( \mathrm{m}\pi)}\,P_{ \nu}^{
\mathrm{m}}(\cosh(\frac{| \mu \Lambda_{b}|}{2}\,\eta))+& \nn\\
 & \,\,\,\,\,\,\,\,\,\,\,\,\,\,\,\,\,\,\,\,\,\,\,\,\,\,\,\,\,\,\,\,\,\,\,\,\,\,\,\,\,\,\,\,\,\,\,\,\,\,\,
 \,\,\,\,\,\,\,\,\,\,\,\,\,\,\,\,\,\,\,\,\,\,\,\,\,\,\,\,\,\,\,\,\,\,\,\,\,\,\,\,\,\,\,\,\,+
 \frac{2}{\pi}\,e^{-i \mathrm{m}
\pi}\,Q_{\nu}^{\mathrm{m}}(\cosh(\frac{| \mu
\Lambda_{b}|}{2}\,\eta))],&
\label{normalized}
\eea
where
$f=\frac{1}{2^{4\mathrm{m}}(\nu+\frac{1}{2})^{2\mathrm{m}}}\,\frac{\Gamma(1-\mathrm{m})\,\Gamma(\frac{1}{2}-\mathrm{m})\,
 \Gamma(1+2\mathrm{m})\,\Gamma(\nu+\mathrm{m}+1)}
 {\Gamma(1-2\mathrm{m})\,\Gamma(1+\mathrm{m})\,\Gamma(\frac{1}{2}+\mathrm{m})\,\Gamma(\nu-\mathrm{m}+1)}$\,\,
 and\, Im$\nu>0$, has the property
\be
\bar{u}_{k}(\eta)\rightarrow\bar{u}_{k}^{(0)}(\eta)\,\,, \,\,
\lambda\rightarrow+\infty .
\label{arrow}
 \ee
This solution refers to the $+\infty$ region; for the $-\infty$
region the above $\bar{u}_{k}$ has to be multiplied by $-i$. For
the derivation of the relation (\ref{arrow}) the following
equations have been used:\\ i)
$P_{\nu-\frac{1}{2}}^{\nu\,\acute{}}(\cosh
x)=\frac{2^{2\nu\,\acute{}}}{\Gamma(1-\nu\,\acute{}\,)}(1-e^{-2x})^{-\nu\,\acute{}}\,e^{-(\nu+\frac{1}{2})x}\,
_{2}F_{1}(\frac{1}{2}-\nu\,\acute{}\,\,,\,\frac{1}{2}+\nu-\nu\,\acute{}\,\,;\,1-2\nu\,\acute{}\,\,;1-e^{-2x})$
\newline (see \cite{abramo}, p. 336)\\
 ii)
$_{2}F_{1}(a,b+r;c;z)=\frac{\Gamma(c)}{\Gamma(a)\,\Gamma(c-a)}\,\{\Gamma(c-a)\,(r
z)^{a-c}\,(1-z)^{c-a-b-r}\,[1+$O$(r^{-1})]\,+$\\
$^{}\,\,\,\,\,\,\,\,\,\,\,\,\,\,\,\,\,\,\,\,\,\,\,\,\,\,\,\,\,\,\,\,\,\,\,\,\,\,\,\,\,\,\,\,\,\,\,\,\,\,\,\,\,\,\,
\,\,\,\,\,\,\,\,\,\,\,\,\,\,\,\,\,\,\,\,\,\,\,\,\,\,\,\,+\Gamma(a)\,(r
z\,e^{-i
\pi})^{-a}\,[1+$O$(r^{-1})]\,\}\,\,,\,\,\,\,\,\,\,|r|\rightarrow\infty$\\
(see \cite{luke}, vol. 1, p. 241) and also the relation between
$Q_{\nu}^{\mathrm{m}}$ and $P_{\nu}^{\mathrm{m}}$,
$P_{\nu}^{\mathrm{-m}}$. Equation (\ref{normalized}) diverges for
$\mathrm{m}\in \mathbf{Z}$, since $\sin(\mathrm{m}\pi)$ vanishes.
The massless field $m=0$ is included in this case. For a negative
cosmological constant $\Lambda_{B}$, an adiabatic vacuum cannot be
defined using the above approximations. Since the solution
(\ref{normalized}) reduces to (\ref{asympto}) in the limit of
large $\lambda$, regardless of the value of $\eta$ (see
\cite{chernikov}), it defines a stable adiabatic vacuum for all
times. Thus, the modes (\ref{normalized}) are positive frequency
with respect to the adiabatic definition, and in this vacuum an
inertial detector registers no particles.

\section*{6 \,\,\,Conclusions}
\hspace{0.8cm} Recently, there has been renewed interest in
cosmological models with extra dimensions. Major effort has been
devoted to branes sitting on fixed positions of the bulk space. In
this work, we have investigated the existence of bulk solutions
arising from a moving domain wall framework. In the beginning we
derived, based on the Israel matching conditions, the equations
governing a moving domain wall in a general class of cosmological
bulk metrics. Subsequently, assuming an ansatz for these metrics
consisting of true time-dependence of the two ``lapse'' functions,
we were in a position to obtain an additional first integral for
the domain wall motion when the matter on the wall is a
cosmological constant. This leads to a condition (unique for only
one scale factor of the $D-2$ part) between the bulk metric
components, necessary for the compatibility of the formalism.
Afterwards, we found for a flat three-universe, a five-dimensional
three-parameter bulk solution admitting the above described wall
motion, when a cosmological constant exists in the whole
spacetime. This solution is shown to be the unique solution under
the conditions assumed. Investigating the properties of this bulk
space, we saw that locally it is simply the well known $dS_{5}$ or
$AdS_{5}$ space. Hence, these spaces admit moving three-universes
whose trajectories are determined by the found solution. The
domain wall formed by the motion of the three-universe corresponds
to the ``physical'' four-dimensional spacetime. This is shown to
be $dS_{4}$. A remarkable fact concerning this three-universe is
that its scale factor does not vanish during its unfolding through
the bulk space, i.e. it avoids collapsing. As the conformal time
goes to infinity, this scale factor reaches a positive minimum,
and the bulk metric has an apparent (coordinate) singularity
there. If another foliation of the $(A)dS_{5}$ space is adopted,
avoiding this singularity, then the fate of the corresponding
three-universe should be examined anew.
\par
For negative bulk cosmological constant, the bulk space being
$AdS_{5}$, the metric could be cast in the form appearing in
\cite{randall, dewolfe}, with the emergence of a bound
gravitational state. Our solution corresponds to a three-universe
that does not reside in a prefixed-position brane. The two
pictures correspond to distinct physical situations and do not
seem to be continuously connected.
\par
As an application of the properties the above bulk metric possess,
we considered the inclusion of a scalar field, neglecting its
back-reaction on the background. The wave equation of the scalar
field is separable and its complete solution has been obtained.
The inclusion of a general domain wall, coupled to the field, can
guarantee the normalizability of the field in the infinite
extra-dimension range. The time-dependent part of the field,
examined in the ``conformal'' gauge, allows, for a negative bulk
cosmological constant, the definition of asymptotically positive
frequency behaved modes, even in the massless limit. For positive
cosmological constant an adiabatic vacuum has been consistently
defined for non-zero mass. Notions of particle creation in the
bulk space can be extracted, though this topic deserves further
investigation.
\par
It seems interesting and also possible to find a 4-parameter
solution containing $\mathcal{C}\neq 0$ and/or a non-flat three
universe. A harder task would be to find any solutions with the
``lapses'' being  functions of both time and the extra coordinate.
The inclusion of two different bulk cosmological constants in the
two sides of the domain wall, relaxing the reflection symmetry
assumption would also be interesting, within string or
supergravity theory, for whatever ansatz of the bulk metric.
Furthermore, a perfect fluid content in the three-universe would
make a moving domain wall model cosmologically more realistic.
Finally, we refer to the work \cite{deffayet}, where a
self-inflationary solution was introduced, which produces a phase
of late accelerated expansion, as indicated by recent supernova
data. This is the result of an intrinsic curvature Ricci scalar,
included in the brane action. It would be desirable if one managed
to embody such a geometric term in the Israel conditions of the
moving formalism.
\section*{Acknowledgements}
We wish to thank Nikolaos Tetradis and Vasilios Zarikas for
helpful discussions.



 \end{document}